\documentclass{doublecol-new}

\usepackage{algorithm}
\usepackage{algorithmic}
\usepackage{epsfig}
\usepackage{multirow}
\usepackage{booktabs}
\usepackage{enumerate}
\usepackage{multicol}
\usepackage{amsmath}
\usepackage{color}
\usepackage{stfloats}
\def\newblock{\hskip .11em plus.33em minus.07em}

\theoremstyle{TH}{

}

\theoremstyle{THrm}{

}

\theoremstyle{THhit}{

}

\usepackage{flushend}
\usepackage[square, comma, sort&compress]{natbib}

\usepackage{url}

\makeatletter
\def\theequation{\arabic{equation}}
\makeatother

 \makeatletter
 \def\url@leostyle{%
   \@ifundefined{selectfont}{\def\UrlFont{\sf}}{\def\UrlFont{\small\ttfamily}}}
 \makeatother
\usepackage{fancyheadings}
\begin{document}%
%%%%%%%%%%%%%%%%%

\thispagestyle{plain}

\setcounter{page}{1}

\LRH{S. Mittal  }
\RRH{}
%\RRH{Accepted in Int. J. of Computer Aided Engineering and Technology, 2014.}

\VOL{x}

\ISSUE{x}

\PUBYEAR{2013}

\title{A Survey of Techniques For Improving Energy Efficiency in Embedded Computing Systems}

 \authorA{Sparsh Mittal}
  \affA{Future Technologies Group\\
   Oak Ridge National Laboratory (ORNL)\\
   Oak Rdge, TN, USA\\
   Email: sparsh0mittal@gmail.com
 }

\begin{abstract}
Recent technological advances have greatly improved the performance and features of embedded systems. With the number of just mobile devices now reaching nearly equal to the population of earth, embedded systems have truly become ubiquitous. These trends, however, have also made the task of managing their power consumption extremely challenging. In recent years, several techniques have been proposed to address this issue. In this paper, we survey the techniques for managing power consumption of embedded systems. We discuss the need of power management and provide a classification of the techniques on several important parameters to highlight their similarities and differences. This paper is intended to help the researchers and application-developers in gaining insights into the working of power management techniques and designing even more efficient high-performance embedded systems of tomorrow.

\end{abstract}

%To address this, several to ensure reliability, performance scaling and . 

%to continue to enhance their features and performance.    

%used and their performance has 

\KEYWORD{Embedded systems, low power design, power management, energy efficiency,  sustainable computing, green computing, classification, review, survey.}

\REF{to this paper should be made as follows: Mittal, S. `A Survey of Techniques For Improving Energy Efficiency in Embedded Computing Systems', Int. J. of Computer Aided Engineering and Technology, 2014.}

 \begin{bio}
 Sparsh Mittal received the B.Tech. degree in electronics and communications engineering from IIT, Roorkee, India and the Ph.D. degree in computer engineering from Iowa State University, Ames, IA, USA. He is currently working as a Post-Doctoral Research Associate at Oak Rige National Laboratory (ORNL), TN, USA. In his B.Tech. degree, he was the graduating topper of the batch and his major project was awarded institute silver medal. He was awarded scholarship and fellowship from IIT Roorkee and Iowa State University. His research interests include cache architectures, embedded systems, energy efficiency and multicore systems. 
 \break
 \end{bio}

\maketitle
  \fancyhf{} % sets both header and footer to nothing
\renewcommand{\headrulewidth}{0pt}
\thispagestyle{fancy}
\lfoot{Accepted in Int. J. of Computer Aided Engineering and Technology, 2014.}
\vfill
\pagebreak

\noindent\parbox{42.4pc}{\leftskip 12.75pc\NINE .\vs{12}}
%\onecolumn
%\noindent\rule{42.5pc}{1.5pt}

%\maketitle

\section{Introduction}
Recent years have witnessed a phenomenal growth in features and applications of embedded systems. Embedded systems such as mobile computing systems now offer integration of video camera, net browser, wireless data modem and phone. Further, it has been estimated that the number of mobile devices has now become almost equal to the population of the world \cite{mobileNumber}. These trends, however, have also presented significant challenges for managing power consumption of embedded systems. Many portable systems have stringent power budgets, such as 1 or 2W, while ultra low-power embedded systems (e.g. wearable systems) have a power budget of  a few milli watts \cite{ghasemzadehultra}. This is in sharp contrast with general-purpose and graphics processors which have the power budget of up to a few hundred watts \cite{GPU590power}. The low power budgets of embedded systems present severe demands for improving their energy efficiency. As an example, a 3G mobile phone receiver requires nearly 40 GOPS (giga operations per second) to handle a 14.4 Mbps channel and doing this in a power budget of 1W would require energy efficiency of 25pJ per operation \cite{dally2008efficient}. Thus, to continue to scale their performance and ensure reliability, longevity and adoption in wide range of applications, power management has become extremely important for embedded systems.

In this paper, we highlight the need of power management in embedded systems and survey several research works which are aimed at improving energy efficiency of embedded systems. To provide insights into the working of these techniques, we classify them on the basis of their key research idea. We believe that this survey will help the researchers and designers in understanding the state-of-the-art in power management of embedded systems and also motivate them to further improve the energy efficiency of embedded systems.     
 
 In a paper of this length, it is not possible to do justice to the broad range of developments in the field of embedded systems and hence, we take the following approach to limit the scope of the paper. We include only those research works that propose methods for improving energy efficiency and also evaluate it. Those works which only evaluate performance improvement are not included although they may also lead to better energy efficiency. We review  application and architectural level techniques and not circuit-level techniques. Since different techniques have been evaluated using different platforms and methodologies, we only focus on their fundamental research idea and do not present the qualitative results.

The remainder of this paper is organized as follows. Section \ref{sec:background} provides a background on embedded systems and also highlights the need of power management. Section \ref{sec:overview} provides an overview and classification of power management techniques in embedded systems. Section \ref{sec:details} discusses some of these techniques in detail. Finally, Section \ref{sec:conclusion} provides concluding remarks and also discusses the future challenges.

% Embedded systems differ from traditional high-performance supercomputers in that power is a first-order constraint for embedded systems; whereas, performance is the major benchmark for supercomputers.

%\textit{Portable, embedded systems place ever-increasing demands on high-performance, low-power microprocessor design.}

%\textit{With Moore’s law supplying billions of transistors on-chip, embedded systems are undergoing a transition from single-core to multicore to exploit this high-transistor density for high performance.  The increase in on-chip transistor density exacerbates power/thermal issues in embedded systems, which necessitates novel hardware/software power/thermal management techniques to meet the ever-increasing high-performance embedded computing demands in an energy-efficient manner.}

%\textit{Mobile computing is about to enter an era of high data rate applications that require the integration of wireless wideband data modems, video cameras, net browsers, and phones into small packages with long battery powered operation times.}

 %and hence, power management is extremely important for these systems.
 
 %Power is a first-order constraint in the design of embedded systems. 
\section{Background} \label{sec:background}
An embedded system is a computing system which is designed for specific control functions and is embedded as part of the complete device which may include hardware and mechanical parts. Thus, in contrast with general-purpose computers (e.g. desktop), an embedded system performs a few   pre-defined tasks, with very specific requirements. Typical examples of embedded systems include MP3
players, smart cameras and cellular phones.  

\subsection{Sources Of Power Consumption}
We briefly review the sources of power consumption in embedded systems and refer the reader to previous work \cite{butts2000static,niu2004reducing} for more details. The power consumption of embedded systems can be broadly divided in two categories, namely dynamic power and static power. The dynamic power ($P_{dyn}$) consumption arises from charging and discharging of the load capacitance, and the short circuit currents. The leakage power ($P_{leak}$) arises due to leakage currents that flow even when the device is inactive.  Thus, we have
  \begin{eqnarray*}
P_{dyn} &=& \alpha C V^2F\\
P_{leak} &=& I_{leak}V
\end{eqnarray*}
Here $\alpha$ shows the switching activity, $F$ shows the operating frequency and $V$ shows the operating voltage. $I_{leak}$ shows the leakage current. With CMOS scaling the leakage power is increasing dramatically \cite{semiconductor2011international}. DVFS based techniques work by reducing dynamic energy, while the techniques which transition the system to low-power aim to reduce leakage energy.     

\subsection{ Importance of power management} 
Power management in embedded systems is important for the following reasons.
\subsubsection{Limited Size and Battery}
For battery-operated mobile embedded systems, energy supply is a crucial limitation. Power consumption leads to heating, which is unacceptable in several domains such as wearable embedded systems. Further, the small size of these systems also limits the amount of heat-dissipation that can be managed.  Smaller power consumption enables use of smaller power supplies and reduced heat-dissipation overhead, which also reduces the cost, weight and area of embedded systems. Thus power management can lead to easier system design.

\subsubsection{Ensuring Longevity}
A 15 degree Celsius rise in temperature increases the device failure rates by up to a factor of two \cite{anderson2003more}. Thus, power dissipation has deleterious effect on reliability of embedded systems and this phenomenon may be crucial for medical devices and mission-critical systems.  
  
\subsubsection{Addressing Inefficiency Arising due to Over-provisioning of Resources }
In embedded systems, idle intervals arise for several reasons, such as pessimistic estimate of worst-case execution time and inherent slack due to relaxed deadline etc.  Despite this, the designers need to provision resources to meet the worst-case performance requirement which leads to energy wastage. Thus, dynamic energy saving techniques can use runtime adaption to trade performance for saving energy. Also, since the embedded systems are typically used for well-defined applications, static techniques can be easily used for per-application tuning of resources.

\subsubsection{Meeting Performance Requirements} 
In recent years, embedded processors  are used to execute resource-intensive applications (e.g. multimedia processing \cite{GupMit08_MIMO,PanMit2009_Baywave,gupta2008guaranteed}) that were originally designed for general-purpose processors. To meet these performance demands, modern embedded processors use many complex features  such as multi-cores, multi-level caches etc. \cite{armprocessor,mipsprocessor,nvidiategra,munir2012high,wang2011dynamic}. These trends have influenced the design of embedded systems to be  optimized for higher performance, instead of lower power consumption.

\subsubsection{Power Challenges Posed by CMOS Scaling}
The advancements in CMOS technology have greatly increased the on-chip transistor densities and speeds. These trends have led to a technology-imposed \textit{utilization wall} which limits the fraction of the chip that can be simultaneously used at full speed within the power budget. Thus, today the processor performance is primarily constrained by energy efficiency and it has been estimated that, if left addressed, power challenges may end future performance scaling \cite{esmaeilzadeh2011dark,venkatesh2010conservation}.  Conversely, techniques for improving energy efficiency can enable the designers to scale performance by executing parallel computations  without violating the power budget.

\subsubsection{Trends in Usage Pattern}
In recent years, mobile computing devices have become the key platform for the mobile convergence applications, e.g. web browsing, imaging, and video streaming. Due to these features, embedded systems have become ubiquitous. Thus, while an individual portable system consumes much less power than a server in the data center, the large user-base of embedded systems makes their total power consumption very high.  

\subsubsection{Enabling Green Computing}
It has been estimated that the ICT (Information and communications technology) contributes nearly 3\% in the overall carbon footprint \cite{smarr2010project}. Thus, power management in embedded systems is also important for achieving the goals of green computing.

\section{Overview}\label{sec:overview}
 Based on their main energy saving approach, we classify the techniques into following categories.
 
 \begin{enumerate}
 \item  DVFS (dynamic voltage and frequency scaling) and  power-aware scheduling based techniques \cite{brock2003dynamic,pillai2001real,srinivasan2003combining,JejGup05_SlackReclaim,luo2001battery,frohlich2011comprehensive,bhatti2010power,marinoni2011platform,qi2010global,yun2011system,kim2008system,zhang2006unified,shin2000power,schmitz2002energy,firouzi2010reliability,mohapatra2003integrated,march2012power,karakehayov2011energy,santos2009power,koedam2011exploiting,mera2010low,kan2010leveraging,chen2010new,chang2011learning,hua2003approaching,hua2003energy,choi2004fine,gheorghita2006application,choi2004dynamic1,yuan2005analysis,wu2003scheduling,shao2007real,mahapatra2005energy,cho2006high,kianzad2005casper,kim2001hybrid,devadas2008interplay,gruian2001lenes,saewong2003practical,quan2001energy,zhu2003scheduling,xian2007energy,park2006low,luo2000power,rakhmatov2003energy,liu2001power,wang2011task,cho2011design,niu2011energy,ma2012feedback,yang2003pareto,yang2002managing,yuan2006grace,de2007trade,chang2008etahm}.

 \item Using low power modes, called power mode management (PMM)  \cite{shukla2001model,frohlich2011comprehensive,shin2000power,huang2010adaptive,muhammad2010data,karakehayov2011energy,li2002mode,marinoni2011platform,devadas2008interplay,hoeller2006hierarchical,cheng2006online,awan2011enhanced,de2007trade}. Note that this is also sometimes referred to as `dynamic power management'. We however use a more specific term `power mode management' to avoid any confusion.
 
 \item Microarchitectural techniques for saving energy in specific components e.g. main memory (\cite{Yang2010OMC,trajkovic2008improving,mamidipaka2003chip}), cache \cite{MitZha12_EnCache,wang2009dynamic,wang2011general,wang2011dynamic,MitZha13_Cashier,reddy2010cache,paul2011dynamically,bournoutian2008miss,hajimiri2011synergistic,mittal2013PhDThesis,petrov2001performance,tseng2009design,mohapatra2003integrated,alipour2011cache,MittalPalettePaper2013,kwak2010selective,niu2004reducing,zhang2003highly,mittalMANAGER2013,chen2007dynamically,KinGup97_FilterCache,albonesi1999selective}, scratchpad memory \cite{steinke2002reducing}, TLB \cite{choi2002low} or making other changes to memory hierarchy e.g. adding extra components \cite{tsai2011energy,benini2000increasing}. 

\item Using unconventional-cores such as DSP or GPUs of FPGAs \cite{timm2010reducing,wang2012energy,gpuDSP2011,wang2011energy,mencer1998hardware,stitt2004energy,biswas2006performance,llamocca2011separable,fowers2012performance,strozek2006efficient}.
 
%\item Application-specific optimizations such as for video processing \cite{he2008energy}, scientific computing \cite{goddeke2012energy} 
  
%\item Other techniques such as   code compression \cite{lekatsas2000code}
\end{enumerate}  
  
  %saving cache energy in embedded multitasking systems (e.g. \cite{wang2011general,reddy2010cache,paul2011dynamically,wang2011dynamic})
  
%///////////////
%Review \cite{rakhmatov2003energy,liu2001power,luo2001battery,frohlich2011comprehensive,kim2008system,firouzi2010reliability}

%Dont review \cite{mamidipaka2003chip,yun2011system}

%multiple components \cite{yun2011system}  exploiting slack \cite{santos2009power}

%,kianzad2005casper,gruian2001lenes

\section{Power Management Techniques}\label{sec:details}
We now discuss some power management techniques in detail. As evident from the  classification in previous section, several techniques can be classified in more than one category (e.g. DVFS and PMM). For sake of brevity, we discuss them in one category only.
\subsection{DVFS and Power-Aware Scheduling based Techniques}
For sake of convenience, we discuss power-aware scheduling techniques along with DVFS since these techniques often make scheduling decisions with a view to use DVFS for saving energy.  

DVFS is a technique for altering the voltage and/or frequency of a computing system based on performance and power requirements. For CMOS circuits, dynamic power is related with voltage and frequency as $P \propto FV^2$ and hence, by reducing the frequency, the voltage at which the circuit needs to be operated for stable operation can also be lowered, which leads to energy saving. Several commercial microprocessors support DVFS technology for saving power, e.g.  AMD PowerNow and and Intel's SpeedStep. The limitation of DVFS is that it harms the performance and hence, it may increase execution time or lead to missed deadlines. Also, DVFS requires programmable clock generator and DC-DC converter which incur energy overhead. Further, voltage transitions may require time  on the order of tens of microseconds (\cite{kim2008system}). Finally, due to increase in leakage energy and trend of using multi-core processor instead of increasing clock frequency, the returns from DVFS are diminishing.

 Hua et al. \cite{hua2003approaching} study the opportunity available for saving energy in embedded systems using DVFS. Since supporting a large number of voltage levels causes overhead (e.g. area and power overhead of voltage regulators, overhead of transitions),  their work explores the optimal number of voltage levels and their values to implement on the multiple-voltage system for achieving energy efficiency. They have shown that systems which provides 3 or 4 voltages are nearly as energy efficient as the ideal system that can vary the voltage arbitrarily.

Kianzad et al. \cite{kianzad2005casper} use genetic algorithm to integrate task scheduling and voltage scaling under a single iterative optimization loop. Their technique  searches the solution space to find an assignment and ordering of tasks on each processing element and generates a schedule such that deadline constraints are met and the power consumption is minimized. Further, their technique distributes the slack proportionately to different tasks and uses DVFS to save energy. They propose techniques  for saving energy in both homogeneous and heterogeneous  multiprocessor embedded systems.

 In several multimedia applications, missing some task deadlines can be acceptable since it remains unnoticed to human visual and auditory system.  Hua et al. \cite{hua2003energy} utilize this fact, along with the information on statistical task execution time to propose techniques to save energy in embedded systems by dynamic voltage scaling. They have proposed two algorithms. The first algorithm ensures achieving highest completion ratio with lowest possible energy consumption. The second algorithm deliberately drops some tasks to create slack for saving additional energy, such that application-specific quality-of-service constraint is fulfilled.  Thus, their algorithms provide opportunity to exercise trade-off between achieving high energy saving and achieving high task completion ratio (i.e. low deadline miss ratio).

 Choi et al. \cite{choi2004fine} propose  a DVFS technique which enables achieving a precise energy-performance trade-off. Their technique makes use of runtime information about the external memory access statistics and chooses the optimal CPU clock frequency and the corresponding minimum voltage level based on the ratio of the on-chip computation time to the off-chip access time. Their technique lowers the CPU frequency in the memory-bound region of a program to keep the  performance degradation to a low value.

Gheorghita et al. \cite{gheorghita2006application} propose a technique for saving energy in embedded systems by utilizing the knowledge of operation modes. As an example, a portable MP3 player may have different application scenarios e.g. actual use for listening, connection with computer etc. Further, the player provides two operation modes viz. mono or stereo. Since mono mode requires less computation power, battery power can be saved by using smaller voltage at the time of low resource usage. This observation can also be used to guide design-time choices to provide specific operation modes for specific use-case scenarios. They also show the application of their technique for both hard and soft real-time systems.

 Saewong et al. \cite{saewong2003practical} propose four DVFS techniques for saving energy in embedded systems. Their first technique selects a single frequency which is used for the entire execution, such that the task-deadline is met and energy is minimized. This technique is suitable for systems where the overhead of DVFS is very high. The third technique finds suitable frequency like the first technique with the difference that it takes decision in the order of priority of  tasks. Hence, if meeting the deadline of a higher priority task requires running other tasks at larger frequency than what is minimally required for meeting their deadlines, extra slack is created which can be used to save extra energy in low-priority tasks by further reducing their frequency. Another technique uses non-linear optimization to find the optimal frequency for every task. This technique, however, has large complexity and hence is unsuitable for on-line use. The fourth technique additionally monitors the actual execution time of tasks and further minimizes the energy consumption of tasks whose execution times are less than the pre-reserved worst-case execution times.

Quan et al. \cite{quan2001energy} propose two DVFS algorithms for saving energy in  real-time embedded systems. The first algorithm finds the minimum constant speed that can be applied throughout the execution of the whole tasks set, such that the processor is  shut down when idle. The second algorithm produces both the constant speed and schedule of variable voltage for minimizing energy. The second algorithm always saves more energy that the first algorithm.

Zhu et al. \cite{zhu2003scheduling} propose a technique for saving energy in  multiprocessor systems. Their technique allows the processors to share the reclaimed slack which arises from a shorter execution time in one of the processors. Using this extra slack, the speed of future tasks can be reduced which results in energy savings. They show that sharing the slack also helps in ensuring that all the deadlines are met. They show the effectiveness of their technique for both tasks with dependence (i.e. precedence constraints) and without dependence.  

Xian et al. \cite{xian2007energy} present a technique for scheduling in multiprocessor systems to save energy. For scheduling periodic real-time tasks, their technique uses earliest deadline first scheduling to ensure meeting the deadlines of all tasks while minimizing energy consumption. Since this problem is NP-hard, they present a polynomial time heuristic method. For this, the problem is solved as a load-balancing problem assuming that unbounded and  continuous range of frequencies are available. In the second step, this solution is modified to account for the maximum available frequency and the bounded discrete frequencies.

Kan et al. \cite{kan2010leveraging} propose a DVFS technique for saving energy in soft real-time embedded systems. They find the optimal frequency for a task assuming availability of continuous range of frequencies. Afterwards, from the actually available frequencies, the closest frequencies which are smaller and larger than the optimal frequency are chosen and fraction of time each of them should be used to meet the deadline is decided. They also suggest that difference in average-case and worst-case execution time gives rise to multiple deadlines and hence, if the first deadline is missed, the algorithm can try to meet the next deadline while conserving extra amount of energy.

Yang et al. \cite{yang2002managing} propose technique for mapping concurrent tasks  onto a heterogeneous multiprocessor platform in an energy efficient manner.  On different processors, the execution time and energy consumption of the tasks are different and hence, their technique finds  different task-ordering and processor-assignment possibilities , and generate a Pareto-optimal set, where every point is better than any other one in at least one way (i.e., either energy efficiency or execution speed). This information is used at runtime to save energy using dynamic voltage scaling and to find a global energy efficient solution.

 Kim et al. \cite{kim2008system} discuss  per-core DVFS technique for saving energy in embedded systems. Scaling CPU frequency slows down the CPU-bound operations but has little effect on memory-bound operations and based on this fact,  the voltage and frequency are reduced during memory-bound intervals of an application. Their algorithm finds suitable voltage/frequency setting in an offline manner. Use of on-chip regulators enables fine-grained voltage transitions using which the memory-bound intervals can be more effectively exploited. They also show that per-core DVFS scheme can better exploit the scaling opportunities in the individual threads and thus provides significant improvement over system-wide DVFS scheme.  Further, they study the effect of voltage transition time, overhead and regulator losses on the benefit obtained from DVFS.

\subsection{Using Power Modes}
In embedded systems, the hardware typically provides a range of operating modes which can be used to save energy. Different modes consume different amount of power and take different time to return back to the normal mode. In general, the modes with lower energy consumption also take the largest time to return to the normal mode  and vice versa. For saving energy while keeping the performance loss bounded, these modes should be judiciously used. Also, while a low-power mode can be used when the system is idle, the system must return to the normal mode for actually servicing a request or performing the task. 
%Several techniques utilize these power modes to save energy.  

Li et al. \cite{li2002mode} propose a method for selecting the power modes for the optimal power management of embedded systems under timing and power constraints. Their method determines the schedule of mode transitions such that the whole system can meet all power and timing constraints.

Hoeller et al. \cite{hoeller2006hierarchical} propose an interface for power management of hardware and software components. They method allows applications to express when certain components are not being used and based on this information, individual components, subsystems or the whole system can be transitioned to low-power modes. This frees the programmer from the task of individually managing the power consumption of each component. 

Huang et al. \cite{huang2010adaptive} propose an energy saving technique which works by  adaptively controlling the power mode of the embedded system according to historical arrivals of tasks. Their technique takes decision regarding when to transition the system to low-power from normal-power mode or vice versa, based on the relative time overhead and energy advantage from mode transition and the consideration of meeting the deadlines of the tasks. 
      
Bhatti et al. \cite{bhatti2010power} present an online framework to integrate DVFS with power-mode management (PMM) scheme to save energy in embedded systems. Their scheme utilizes conventional DVFS and PMM schemes and uses machine-learning approach to adapt at runtime to the best-performing policy for any given workload. To save energy, DVFS policy makes use of dynamic slack while PMM makes use of idle time intervals. They have shown that their technique achieves energy savings comparable to the best-performing policy at any time. Also, their framework allows use of existing as well as new DVFS and PMM schemes.
 
Niu et al. \cite{niu2004reducing} propose a technique to save both leakage and dynamic energy in embedded systems by integrating DVFS and PPM. In the case when processor is active, their technique chooses a processor speed such that the dynamic and leakage power consumption are balanced. Further, in the case when the processor is idle, the coming tasks are delayed as much as possible, such that their deadlines are not missed and scattered, short inter-task idle intervals are merged in a few large idle intervals. Large idle intervals lead to reduced mode transition overhead, since the processor can stay in either idle or active state continuously for longer time. 
   
Kim et al.    \cite{kim2001hybrid} study the trade-off between voltage scaling and dynamic power-mode management. Under the assumption that voltage scaling does not reduce energy consumption in peripheral devices, voltage scaling increases the execution time and thus the leakage energy consumption of peripheral devices is increased and opportunity to transition them to low-power mode is reduced. Towards this, they propose a technique which exploits task-slack  by partitioning the task execution into several intervals and shuts down the unneeded peripheral device on a per-interval  basis. 
   
Shin et al. \cite{shin2000power} propose a technique for saving energy by integrating DVFS and PMM. They note that in real-time embedded systems, idle intervals can arise due to either inherent slacks, need to maintain priorities or early completion of tasks than their worst case estimates. To exploit inherent slack for saving energy, they use an offline DVFS approach. Further, they use an online approach which evaluates both DVFS and PMM to find the best way to exploit the remaining two kinds of idle intervals for saving energy.        
      
Cheng et al. \cite{cheng2006online} propose an online technique for performing energy-aware I/O scheduling  for hard real-time systems. Their technique utilizes device slack to perform power mode transitions to save energy, while maintaining temporal correctness. Their technique performs inter-task scheduling and not intra-task scheduling, since it may lead to missed deadlines which may have severe consequences in hard real-time systems. The decision to transition are taken based on break-even time calculation, which shows the minimum time the device needs to be idle for the mode transition to provide positive energy savings.

Awan et al. \cite{awan2011enhanced} propose an approach for saving energy in embedded systems using multiple low-power modes. Their technique computes the break-even time for each mode using offline analysis. Further, since early completion of high-priority task creates slack, their technique accumulates this task and uses it to save extra leakage energy in lower priority tasks by allowing the device to stay in low-power mode for longer time.

\subsection{Saving Energy in Specific Components}
Several researchers propose microarchitectural techniques for saving energy in specific components of embedded systems. These techniques leverage application properties or variation in workload to dynamically reconfigure the component of the system to save energy.

Yang et al. \cite{Yang2010OMC} discuss a technique for saving main memory energy in embedded systems. Their technique uses software-based RAM compression to  increase the effective size of the memory. The memory compression is used only for those applications which may gain benefit in performance or energy  from the compression. For such applications, compression of memory data and swapped-out pages is performed in an online manner, thus dynamically adjusting the size of the compressed RAM area.

 Trajkovic et al. \cite{trajkovic2008improving} propose a buffering based technique for saving energy in low-power embedded systems. Their technique is based on the observation that since the DRAMs allow the row to be left `on' after a memory access; if in a synchronous DRAM (SDRAM), two memory access (i.e. read/write) operation are done in a same activate-precharge cycle, then the overhead of activation and precharging can be avoided. Using this observation, their technique prefetches additional cache blocks on read accesses and  combines multiple blocks (which are to be written to the same DRAM row) in write accesses.  Their technique uses small storage structures  to store the extra prefetched lines and to buffer the writes to the same DRAM row. By adapting the above mentioned write-combining and prefetching schemes for each application, their technique reduces the memory power consumption.

Reddy et al. \cite{reddy2010cache} present an approach for saving cache energy in multitasking embedded systems. Their algorithm selects best cache partitioning for different running applications in offline manner and uses this information to allocate cache at runtime. Their algorithm also reduces inter-task interference in a preemptive multitasking environment. 

Tsai et al. \cite{tsai2011energy} propose a technique for saving energy in embedded processors by using a memory structure called ``Trace Reuse cache'' (TRC). The TRC is used at the same level of memory hierarchy as conventional instruction cache. TRC reuses the retired instructions from the pipeline back-end of a processor and efficiently delivers the instructions in the form of traces. Thus, TRC enables the processor to achieve a higher instruction rate, which leads to improvement in both performance and energy efficiency.     

 Hajimiri et al.\cite{hajimiri2011synergistic} integrate cache reconfiguration and code compression to improve both performance and energy efficiency of embedded systems.  For a single-level cache hierarchy, their technique performs exhaustive exploration of the cache design space by varying different parameters such as line size, associativity and total size; and simulating each one of the resultant configuration. Based on the results, the configuration with minimum energy can be selected. The code compression scheme integrates synergistically with cache reconfiguration, since code compression improves performance by reducing memory traffic and bandwidth usage and thus it partially offsets the performance loss resulting from cache reconfiguration.

It is well-known that there exists a large intra- and inter-program variation in cache requirement of different applications. Using this, the cache can be dynamically reconfigured for each program or program phase and the unused cache is turned off to save energy. Based on this idea, Albonesi \cite{albonesi1999selective} proposes selective-ways approach where some of the ways of the cache are turned off to save energy, such that performance degradation remains bounded.

Zhang et al. \cite{zhang2003highly} propose a highly-configurable cache architecture for facilitating dynamic reconfiguration to save energy in embedded systems. Their cache architecture contains four separate banks that can operate as four separate ways. By concatenating these ways, the associativity of the cache can be changed to either 1, 2 or 4. Also, if desired, some ways can be shut down. Further, by configuring the fetch unit to fetch different size of cache lines, the line size (block size) of cache can also be altered.

Several researchers use this architecture to save cache energy. For single-core systems, Wang et al. \cite{wang2009dynamic} profile several possible configurations of L1 data cache, L1 instruction cache and unified L2 cache in offline manner. At runtime, different possible combinations of two-level cache hierarchy are explored to find the most energy efficient configuration. For multi-core systems with private L1 caches (data and instruction) and shared L2 cache, Wang et al. \cite{wang2011dynamic} propose using static profiling for exploring different possible combinations and dynamically reconfiguring L1 cache and partitioning L2 cache for saving energy. Similarly, Rawlins et al. \cite{rawlins2011cpact} discuss their technique for saving cache energy in heterogeneous dual-core systems by tuning the size of L1 cache, while addressing the issues presented by multicore operation such as core-interactions, data coherence etc.

Kin et al. \cite{KinGup97_FilterCache} propose a technique for saving energy in embedded systems by filtering access to cache. Their technique places a small filter cache  in front of the conventional L1 cache. Based on temporal locality of data access, most of the accesses are served from the data present in the filter cache and hence the number of L1 accesses is reduced which saves dynamic energy. The trade-off involved in their technique is that for achieving a reasonably high hit-rate in the filter cache, large filter cache is required  which, in turn, increases its access time and energy consumption.

Some researchers have used scratchpad memory for saving energy in embedded systems. Scratchpad memory refers to on-chip SRAM that is mapped into an address space disjoint from the off-chip memory but connected to the same address and data buses. Compared to off-chip memory, both cache and scratchpad allow much faster access. The main difference between the cache and scratchpad is that the cache access may lead to either hit or miss, while the scratchpad guarantees a single-cycle access time. Steinke et al. \cite{steinke2002reducing} propose a technique for saving energy in embedded systems by utilizing scratchpad memory.  At compile time their technique inserts copy functions into the application to copy a set of basic blocks which are frequently executed.   Afterwards, at certain predetermined points in the execution of the program, their technique copies part of the program in the scratchpad and then executes the program from the scratchpad itself. This reduces the access to cache and thus leads to saving of energy. The tradeoff in their approach is that copying instructions into scratchpad consumes more time and energy than a hardware-controlled cache fill.

Benini et al. \cite{benini2000increasing} discuss the design of application-specific memory to save energy in embedded systems. Their technique works by mapping most frequently accessed locations onto a small memory that can be placed on-chip.
Since such a memory is very close to the processor, access to it causes much smaller overhead than accessing large  background memory. Their technique does not use cache as the local memory, since caches incur the overhead of tag comparison. Instead they use a local memory and create application-specific decoding logic which is automatically synthesized at design time based on profiling the embedded application on the processor. Compared to scratchpad, their local memory and decoding logic ensure that most frequently accessed data are stored in a small number of contiguous memory addresses.

Bournoutian et al. \cite{bournoutian2008miss} present a technique for saving energy in embedded systems by reducing L1 cache misses. Their technique employs a small list to hold the set-number from which this line was evicted. On any cache eviction, this list is searched and if an entry with same set number is found, it indicates that the set is in probable state of thrashing. For such sets, a flag is turned on. When a cache miss is observed in sets with the flag turned on, a complementary set is also searched and if the data is found there, the cache access leads to hit. Thus, their technique effectively doubles the associativity of heavily used sets. This reduces the L1 cache misses which in turn reduces the overall execution time and power consumption by avoiding secondary memory accesses.

Mohapatra et al. \cite{mohapatra2003integrated} present a power management technique which integrates microarchitectural level, OS level and middleware level schemes for saving energy in mobile handheld devices. At microarchitectural level, they use static profiling to explore various cache configurations to find the energy-optimal configuration. At OS level, they use DVFS to save energy by exploiting the slack. At middleware level, they use network traffic regulation and admission control policies to save energy in network interface. They also study the interaction of these schemes to further optimize them for saving energy.

\subsection{Using Unconventional Cores}
As discussed before, the performance of modern computing systems is primarily shaped by power concerns. In such a regime, ``unconventional'' cores (or platforms) such as GPUs, FPGAs, ASICs and DSPs etc. \cite{mittal_fpga,keckler2011gpus} hold a great promise for improving application performance and energy efficiency. For this reason, several researchers have used unconventional cores for power management of embedded systems.

Mu et al. \cite{gpuDSP2011} compare the energy efficiency of GPUs with that of DSPs for  
high performance embedded computing (HPEC) benchmark suite which includes a broad range of signal processing applications. They have observed that although GPU provides at least an order of magnitude better performance than the DSP, its energy efficiency (measured in performance per watt) is inferior to that of the DSP. 

Mencer et al. \cite{mencer1998hardware} compare the energy efficiency of FPGAs with that of DSPs for IDEA (International Data Encryption Algorithm). They have observed that the FPGA provides an order of magnitude better energy efficiency than the DSPs.   

Timm et al. \cite{timm2010reducing} compare the performance and energy efficiency of CPU with that of GPUs for several applications, such as matrix multiplication and FFT etc. That have observed that GPU offers significant performance advantage over CPU and hence, despite consuming larger peak power, it outperforms CPU in energy efficiency. They also note that the advantage of GPU reduces for applications which do not provide substantial parallelism.
 
Ou et al. \cite{ou2003performance} propose a technique for energy-efficient mapping of embedded signal-processing applications on FPGA. They use dynamic-programming based approach for mapping beamforming applications which are used in air-borne or sea-borne vehicles. They have shown that compared to a greedy algorithm, their technique provides much larger energy savings.

Wang et al. \cite{wang2012energy} compare the performance and energy benefits of utilizing the integrated GPU and DSP cores to offload or share the compute-intensive
tasks of CPU. They test three mobile computing systems viz. TI's OMAP3530, Qualcomn's Snapdragon S2, and Nvidia's Tegra 2. All these systems integrate both CPU and GPU. Further, TI's OMAP also integrates DSP and exposes it for programming by the user. They test three applications viz. FFT, matrix multiplication and 2D stencil. They have observed that by effectively using GPU and DSP along with CPU, significant improvement in performance and energy efficiency can be achieved.
 
 Stitt et al. \cite{stitt2004energy} propose an approach for moving critical software loops to reconfigurable hardware for saving energy. Typical benchmark programs spend a large fraction (e.g. 80\%) of their time in a small portion of code. Thus, by Amdahl's law, to improve overall performance, this portion can be implemented on a application-specific or reconfigurable hardware which provides much better performance than the software. They demonstrate their approach by using ASIC and FPGA and observe large energy savings over a software-only implementation.
    
Llamocca et al.  \cite{llamocca2011separable} compare the performance and energy efficiency of GPU and FPGA for 2D FIR (finite-impulse response) filter. This program finds application in video processing. They have observed that although FPGA provides lower performance than the GPU, it outperforms GPU on the metric of energy efficiency. The high performance of GPU is due to higher frequency and its ability to exploit massive parallelization present in the algorithm.

Fowers et al. \cite{fowers2012performance} evaluate sliding window program on multi-core CPU,  FPGA and GPU. This program has applications in digital signal processing. They have observed that FPGA provides an order of magnitude better performance while using an order of magnitude less energy compared to both CPU and GPU. These results demonstrate the utility of FPGAs for implementation of embedded system performing high-definition video processing. 

%\cite{strozek2006efficient} 

  %providing the only realistic embedded system implementation for high-definition video

  %compare the energy efficiency of GPUs with that of FPGAs and multi-core CPUs for convolution problem which has applications in digital signal processing. They observe that for very small signal sizes, CPUs are most energy efficient. However as the signal size increases, the energy efficiency of GPUs and FPGAs increase and for very large signal sizes, FPGAs outperform GPUs in energy efficiency.   

\section{Concluding Remarks}\label{sec:conclusion}
% Power management is a central issue in the design of embedded systems. 

The next generation mobile computing systems will possess capabilities for high-speed video processing and communication which will require at least an order of magnitude better energy efficiency than what is available in state-of-the-art systems. This clearly highlights the need of power management in embedded systems. To cope with these challenges, power management is necessary at all levels, viz. chip-design level, microarchitectural level, application level and system level. 
 
In this paper, we reviewed several power management techniques for embedded systems and classified them based on their key research idea. It is hoped that by providing insights into the working of power management techniques, this paper would help the researchers in addressing the challenges of power consumption and architecting highly-energy efficient embedded systems of tomorrow.

%highlight the need of power management in embedded systems and survey several research works which are aimed at improving energy efficiency of embedded systems. To provide insights into the working of these techniques, we classify them on the basis of their key research idea. We believe that this survey will help the researchers and designers in understanding the state-of-the-art in power management of embedded systems and also motivate them to further improve their energy efficiency.     

\bibliographystyle{IEEETR}
\bibliography{References}

%AMSALPHA --shortcut. ACM-like IEEE. AGU-> square
%, SIAM, SPIEBIB, OSA- like IEEE

% \def\notesname{Note}
% 
% \theendnotes

\end{document}